\documentstyle{laa}

\begin{document}
\newcommand{\reff}{\noindent\hangindent=3em\hangafter=1}
\newcommand{\rb}{\right]}
\newcommand{\lb}{\left[}
\newcommand{\per}{$^{-1}$}
\newcommand{\mc}{\multicolumn}
\newcommand{\degree}{$^{\circ}$}
\newcommand{\SM}{M$_{\odot}$}
\newcommand{\citebare}[1]{{citename{#1} citeyear{#1}}}
\newcommand{\amucite}[1]{{(citename{#1} citeyear{#1})}}

\thesaurus{11 (11.03.1; 11.03.4 Abell 2717; 12.03.3; 12.04.1; 12.07.1;
 13.25.2)}

\title{An X-ray and Radio Study of the Cluster A2717}

\author{
H.~Liang\inst{1}
\and M.~Pierre\inst{1}
\and A.~Unewisse\inst{2,3}
\and R.W.~Hunstead\inst{2}}

\offprints{H.~Liang}

\institute{CEA/DSM/DAPNIA CE Saclay, Service d'Astrophysique, F-91191 Gif sur Yvette, France
\and
 School of Physics, The University of Sydney, NSW 2006, Australia
\and
 Electronic and Surveillance Research Lab., DSTO, Salisbury, SA 5108, Australia}

\date{Received ...}

\maketitle

\begin{abstract}
We present an X-ray, radio and optical study of the cluster
A2717. The central D galaxy is associated with a Wide-Angled-Tailed
(WAT) radio source. A {\em Rosat} PSPC observation of the cluster shows that
the cluster has a well constrained temperature of
$1.9^{+0.3}_{-0.2}\times 10^{7}$ K. The pressure of the intracluster
medium was found to be comparable to the mininum pressure of the radio
source suggesting that the tails may in fact be in equipartition with
the surrounding hot gas.

\end{abstract}

\input psfig
\input epsf

\section {Introduction}
The intracluster media of clusters of galaxies can be probed directly by
their X-ray emission and at the same time indirectly through the
morphologies of the radio sources associated with cluster members.
The physical nature of radio galaxies is reflected in the structure of
their emission. Studies of radio galaxies have led to a better
understanding of both radio emission processes and galaxy
environments. One class of interesting radio sources is the
Wide-Angle-Tailed (WAT) sources. Originally defined by Owen \& Rudnick
(1976), WATs are characterised by a radio structure composed of twin
jets separated by a wide opening angle and are associated with the
dominant member of a cluster of galaxies. They are radio sources with
intermediate radio luminosities in the range $10^{42}$ to
$10^{43}$\,erg\,s\per (O'Donoghue 1990), comparable to the
Fanaroff-Riley (1974) break between the FR class I and II.  WATs are
ideally suited for the study of galaxy/environment interactions.  They
are known to favour X-ray poor environments and almost never found in
regions with cooling flows (Norman {\em et al.} 1988; Burns
1990; Zhao {\em et al.} 1990) despite being associated with galaxies
most likely to be found in cooling flow regions. This implies that the
radio emission and cluster gas are associated in some way.

The southern galaxy cluster A2717 has a central D galaxy which is associated with a
WAT source. A2717 has been studied
extensively in the optical (eg. Colless {\em et al.} 1987, 1989). The
known properties of the cluster and its central D galaxy are
summarised in Table~\ref{t:2717cltab} and~\ref{t:2717galtab}.
\begin{table}
\caption{Optical data for the cluster A2717}
\label{t:2717cltab}

\begin{center}
\begin{tabular}{ll} \hline
\mc{2}{c}{} \\[-4mm]
Centre (J2000)$^1$ & RA = $~~00~03~12.4$ \\
		& Dec = $-35~57~48$ \\
Bautz-Morgan class$^2$ & I-II \\
Richness class$^2$ & 1 \\
Redshift$^3$, $z_{cl}$   & $0.0492 \pm 0.0003$ \\
Velocity dispersion$^3$, $\sigma_{los}$ & $547^{+92}_{-72} $\,km\,s\per \\
Schechter characteristic mag$^4$, M$^{*}$ & $-19.55 \pm 0.25$ \\
\hline
\end{tabular}
\end{center}
{\em $^1$ Centroid derived from COSMOS catalogue;
$^2$ Abell {\em et al.} 1989;
$^3$ Colless \& Hewett 1987;
$^4$ Colless 1989.}
\end{table}

\begin{table}
\caption{Optical data for the D galaxy of A2717}
\label{t:2717galtab}

\begin{center}
\begin{tabular}{ll} \hline
\mc{2}{c}{}\\[-4mm]
Position (J2000)$^1$ & RA = $00~03~12.7~~$ \\ 
& Dec = $-35~56~13~~$ \\
Redshift$^2$, $z_{gal}$ & $0.0497 \pm 0.0001$\\ 
Stellar dispersion$^2$, $\sigma_{s} $ & 258$\pm$ 42 km\,s\per \\ [2mm] 
Semi-major axis$^3$ & $36.1^{\prime\prime}$ \\ 
Semi-minor axis$^3$ & $34.4^{\prime\prime}$ \\ 
Position angle$^3$ & $136^{\circ}$ \\ 
B$_{\rm J}$ magnitude$^{3}$ & $14.7$ \\
\hline
\end{tabular}
\end{center}
{\em $^1$ Position measured from an AAT CCD image by R.Hunstead;
$^2$ Colless \& Hewett 1987.
$^3$ Results from COSMOS.}
\end{table}

We present here radio and X-ray observations of the cluster.  In the
radio, it was originally imaged with the Molonglo Observatory
Synthesis Telescope (MOST) at 843 MHz in September 1988 and the host
galaxy of the WAT was found to be the dominant galaxy of the
cluster. In June and October of 1989 two more images were made with
the VLA in scaled arrays at 20\,cm in C/D and 6\,cm B/C in order to
resolve the radio structure. Then, in 1992, a higher resolution image
was obtained with the Australia Telescope (AT) at 6cm. Subsequently, a
{\em Rosat} PSPC pointed X-ray observation was obtained for the
cluster in 1993.

In section
\ref{s:xray}, we present the {\em Rosat} PSPC data and the results of
the analysis of both the flux densities of the discrete sources and
the cluster X-ray suface brightness and temperature. In
section~\ref{s:mass}, we calculate the total mass, gas mass and
compare the X-ray and optical results. In section
\ref{s:radio} we present the radio images at 1.4 and 4.9\,GHz. In
section \ref{s:press}, both limits on the radiation pressures due to
the synchroton emission and the X-ray thermal pressure are calculated.
A value of $H_0 = 50$\,km\,s\per Mpc\per and $q_{0}=0.5$ are used 
throughout the paper.

\section {X-ray Observations} \label{s:xray}
A2717 was observed with the {\em Rosat} PSPC in June 1993 with a total
integration time of 13.7 ksec. The PSPC has a field of view of
$2^{\circ}$ and is sensitive to photons with energy between 0.1 and
2.4 keV. The pointing centre was 00 03 19.2 $-$35 57 00 (J2000.0).
The resolution of the PSPC is $\sim 20^{''}$ at the centre of the
detector at $\sim 1.2$ keV. The resolution degrades with decreasing
energies and increasing distance from the detector centre.

\subsection {Analysis of the X-ray data} 
The {\em Rosat} PSPC data was analysed using the EXSAS package
(Zimmermann {\em et al.} 1994) in MIDAS.
Discrete sources were detected first and subtracted from the data, so that 
they can be removed from the intracluster gas emission.

\subsubsection{Discrete sources}

The discrete sources were detected from a broad band image (including
photons of energy from $\sim 0.11$ to $\sim 2.35$ keV) of $15^{''}$
pixels excluding the regions affected by the support ribs of the
detector. The sources were detected using a sliding window source
detection technique whereby a $3\times 3$ pixels window was slid over
the image. The source counts were obtained from this sliding window
and background counts were determined from the surrounding 16
pixels. This method is best suited for the detection of point or
slightly extended sources.  Finally a maximum likelihood method
(Cruddace {\em et al.} 1988) was applied to the sources detected to
ascertain the reality and the extension of the
sources. Table~\ref{t:ptxray} lists all the sources found with maximum
likelihood parameter $> 10$. These sources were also scanned by eye
and found to be consistent with real sources. None of the sources in
Table~\ref{t:ptxray} can be considered as extended. The positions and
count rates for the sources are listed in Table~\ref{t:ptxray}.
Correlations of the X-ray point sources with the SIMBAD and NED
on-line catalogues were made, but no identification was found.  A
search through the {\em COSMOS} source list with an error radius of
$20^{''}$ produced a number of identifications.  The optical position
and magnitudes of the identified objects within the error radius are
tabulated in Table~\ref{t:ptopt}.

\begin{table}
\caption{Discrete X-ray sources in the field of A2717}
\label{t:ptxray}
\vspace{1.5mm}
\begin{center}
\begin{tabular}{cccc}   \hline 
\multicolumn{1}{c}{source} & \multicolumn{1}{c}{$\alpha$ (2000.0)} &
 \multicolumn{1}{c}{$\delta$ (2000.0)} &
\multicolumn{1}{c}{$f_{x}$} \\
\multicolumn{1}{c}{} & \multicolumn{1}{l}{~h~~m~~~s} &
 \multicolumn{1}{l}{~~~$^\circ$~~~$'$~~~$''$} &
\multicolumn{1}{l}{counts/s} \\
\hline
1 & 00 04 08.31 & $-35$ 54 48.7 & 0.00750 \\
3 & 00 03 59.68 & $-35$ 56 48.4 & 0.00591 \\
4 & 00 03 33.94 & $-36$ 07 34.2 & 0.01081 \\
5 & 00 03 49.99 & $-36$ 07 45.5 & 0.00677 \\
6 & 00 03 29.57 & $-36$ 11 46.3 & 0.00864 \\
7 & 00 02 08.93 & $-35$ 41 58.0 & 0.01604 \\
8 & 00 02 41.79 & $-35$ 55 27.9 & 0.00420 \\
9 & 00 01 49.85 & $-35$ 57 00.7 & 0.01212 \\
10 & 00 03 13.27 & $-36$ 01 46.3 & 0.00529 \\
11 & 00 04 11.97 & $-36$ 02 26.6 & 0.00434 \\
12 & 00 04 39.22 & $-36$ 02 37.6 & 0.00874 \\
13 & 00 02 10.62 & $-36$ 04 20.6 & 0.00495 \\
14 & 00 04 12.55 & $-36$ 06 09.5 & 0.00318 \\
15 & 00 04 18.88 & $-36$ 10 38.9 & 0.00811 \\
16 & 00 04 08.32 & $-35$ 32 19.2 & 0.01389 \\
17 & 00 01 01.24 & $-35$ 46 38.8 & 0.03869 \\
18 & 00 00 52.40 & $-35$ 58 33.7 & 0.01708 \\
19 & 00 02 22.82 & $-36$ 29 18.5 & 0.02162 \\
20 & 00 01 41.08 & $-36$ 03 04.3 & 0.01124 \\
21 & 00 01 05.12 & $-36$ 17 52.1 & 0.01965 \\
\hline
\end{tabular}
\end{center}
{\em Explanation of columns: col.(1) gives the source sequence number; col.(2) \& (3) gives the right ascension and declination of the X-ray source positions in J2000.0 coordinates; col.(4) gives the X-ray flux in count rates.}
\end{table}

\begin{table}
\caption{Optical I.D. of the discrete X-ray sources}
\label{t:ptopt}
\vspace{1.5mm}
\begin{center}
\begin{tabular}{cccccc}   \hline
\multicolumn{1}{c}{source} & \multicolumn{1}{c}{$\alpha$ (2000.0)} &
 \multicolumn{1}{c}{$\delta$ (2000.0)} & \multicolumn{1}{c}{$\Delta r$}
& \multicolumn{1}{c}{$B_{j}$} & \multicolumn{1}{c}{class} \\
\multicolumn{1}{c}{} & \multicolumn{1}{l}{~h~~m~~~s} &
 \multicolumn{1}{l}{~~~$^\circ$~~~$'$~~~$''$} & \multicolumn{1}{l}{~~~$^{''}$}
& \multicolumn{1}{l}{} & \multicolumn{1}{l}{} \\
\hline
1 & 00 04 08.34 & $-35$ 54 50.4 & 1.8 & 21.3 &  \\
  & 00 04 08.01 & $-35$ 54 41.4 & 8.2 & 21.9 & \\
3 & 00 03 59.09 & $-35$ 57 05.1 & 18.2 & 22.0 & \\
4 & 00 03 34.29 & $-36$ 07 44.8 & 11.4 & 18.7 & s \\
5 & 00 03 50.65 & $-36$ 07 57.5 & 14.4 & 15.0 & s \\
6 & 00 03 29.89 & $-36$ 11 46.9 & 4.0 & 21.1 & g \\
  & 00 03 29.97 & $-36$ 12 03.6 & 17.4 & 19.6 & s \\
7 & 00 02 09.28 & $-35$ 42 4.80 & 8.1 & 19.1 & s \\
  & 00 02 08.11 & $-35$ 42 11.3 & 16.6 & 22.0 & \\
8 & 00 02 41.63 & $-35$ 55 37.4 & 9.7 & 21.5 & \\
  & 00 02 43.05 & $-35$ 55 37.7 & 18.2 & 22.5 & \\
9 & 00 01 49.51 & $-35$ 57 16.7 & 16.5 & 18.0 & s \\
10 & 00 03 13.39 & $-36$ 01 53.4 & 7.3 & 21.6 & \\
   & 00 03 12.31 & $-36$ 01 35.1 & 16.2 & 17.5 & s \\
11 & 00 04 12.20 & $-36$ 02 19.2 & 7.9 & 19.8 & g \\
   & 00 04 11.79 & $-36$ 02 07.8 & 18.8 & 17.2 & s \\
   & 00 04 12.20 & $-36$ 02 14.9 & 11.7 & 18.3 & s \\
12 & 00 04 40.46 & $-36$ 02 43.0 & 16.0 & 18.6 & s \\
13 & 00 02 10.60 & $-36$ 04 24.1 & 3.5 & 21.7 & \\
15 & 00 04 18.59 & $-36$ 10 44.6 & 6.7 & 22.8 & \\
17 & 00 01 01.66 & $-35$ 46 25.8 & 14.0 & 21.8 & \\
18 & 00 00 50.99 & $-35$ 58 36.5 & 17.4 & 22.9 &  \\
20 & 00 01 41.24 & $-36$ 03 10.6 & 6.6 & 18.3 & s \\
   & 00 01 42.29 & $-36$ 03 12.6 &  17.0 & 20.4 & g \\
21 & 00 01 05.45 & $-36$ 17 52.9 & 4.1 & 11.6 & s \\
\hline
\end{tabular}
\end{center}
{\em Explanation of columns: Col.(1) gives the source sequence number;
col.(2) \& (3) gives the optical position of all the I.D.s found
within the error radius of the X-ray position; col.(3) gives the
distance in arcseconds between the X-ray position and the optical
objects found with COSMOS; col.(4) gives the optical magnitude;
col.(5) gives the classification of the objects given by COSMOS
where it was able to make a clear distinction between a star (``s'')
and a galaxy (``g'').}
\end{table}

In order to separate the contributions from the discrete sources and
that of the cluster, we subtract the discrete sources found above,
from the photon events file before making images of the field in a
soft (0.1-0.4 keV) and a hard (0.4-2 kev) band (Figure 1).
The peak of the hard image was identified with the central D-galaxy
within uncertainty limits. The hard image peaks at 00 03 12.5
$-$35 56 02 which is $11^{''}$ north of the optical position of the
D-galaxy. The majority of the discrete sources falling within the
central $\sim 45^{'}$ panel of the detector are systematically found
to be between $5^{''}$ -- $15^{''}$ north of their optical counter part,
suggesting a possible systematic error of $\sim 10^{''}$ in the {\em Rosat}
positions. A similar discrepancy of $11^{''}\times 16^{''}$ between
{\em Rosat} PSPC positions and optical positions was found by Pinkney {\em
et al.} (1994). The peak of the soft image was displaced from that of
the hard image and the D-galaxy, but coincides with a faint galaxy
(see Figure 1). This displacement is consistent with the
galaxy being an X-ray source with a spectrum considerably softer than
the cluster ({\em e.g.} an AGN). This soft source was not detected
with the above detection algorithm in the broad-band because of the
soft source spectrum and the proximity of the source to the centre of
the cluster emission which was predominately composed of hard
photons. The position of the soft source was 00 03 16.2 $-$35 56 12
(J2000.0).

There were 2 Abell clusters, A4074 and S1170, and a galaxy group ESO
349-26 that were identified with X-ray sources near the edge of the
detector. These sources were not detected using the formal detecting
methods described above, owing to the proximity of the sources to the
edge of the detector.

\begin{figure}
\label{f:sho}
\epsfxsize 250pt \epsfbox{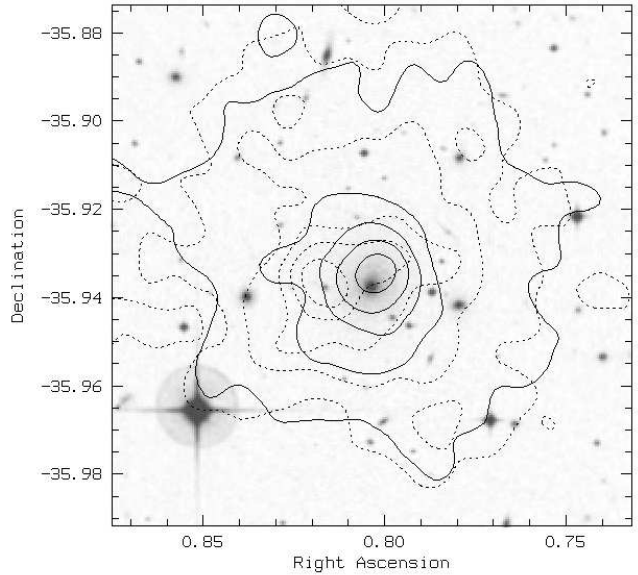}
\caption{The X-ray soft and hard contour images are superimposed on the 
digitised sky-survey (SERC J plates) image of the field. The axes are
in decimal degrees. The X-ray maps have been smoothed with a Gaussian
filter of FWHM $25^{''}\times 25^{''}$.  The hard-band image
(0.4--2keV) is represented by solid contours of levels (1.2, 4.0, 6.8,
10.8) $\times 10^{-6}$\,counts\,s\per\,arcsec\,$^{-2}$; the soft-band
image (0.1--0.4keV) is represented by the dotted contours of levels
(0.8, 1.6, 2.4, 3.2) $\times 10^{-6}$\,counts\,s\per\,arcsec\,$^{-2}$;
and the optical image is in grey-scales. The central D-galaxy is
identified with the peak of the hard image, whereas the peak of the
soft image is identified with a faint galaxy and is displaced from the
peak of the hard image.}
\end{figure}

\subsubsection{The Cluster}

The {\em Rosat} PSPC has limited sensitivity in the high energy part of the
X-ray spectra, namely $>2.4$ keV, thus it can only constrain the
temperature of relatively cold clusters with high precision.

The total cluster spectrum was obtained by including all the photons
within a radius of $500^{''}$ (i.e. $0.65$ Mpc) but excluding all the
regions contaminated by the discrete sources, i.e. a circular area
with a radius of the FWHM of the PSF at the position of the discrete
source. The background spectrum was calculated from an annulus centred
at the cluster centre and between a radius of $1750^{''}$ and
$2250^{''}$.

The X-ray total flux and temperature were derived by fitting a
Raymond-Smith spectrum with galactic absorption to the spectrum, using
a program kindly provided by M. Arnaud.  A $\chi^{2}$ fit was made by
allowing the X-ray emission measure, temperature and the neutral
hydrogen column density to vary but fixing the abundance.  A change in
abundance between 0.3 to 0.5 only produced a $0.1$ keV change in the
best fit temperature. For an abundance of 0.3, the best fit found for
the temperature is shown in Table~\ref{t:x2717} which is rather low
for a cluster. The neutral hydrogen density was found to be
$N_{H}=(1.56\pm 0.13) \times 10^{20}$ cm$^{2}$ as compared to
$N_{H}=1.12\times10^{20}$ cm$^{2}$ from the radio surveys (Stark {\em
et al.} 1992). The X-ray temperature was very well constrained and
relatively independent of $N_{H}$.

\begin{table}
\caption{Results from X-ray analysis of A2717}
\label{t:x2717}

\begin{center}
\begin{tabular}{ll} \hline
\mc{2}{c}{}\\[-4mm]
Centre (J2000.0) & RA = $00~03~12.5~~$ \\
                 & Dec = $-35~56~02~~$ \\
$T_{g}$  & $1.9^{+0.3}_{-0.2}\times 10^{7}$ K  \\
$N_{H}$    & $(1.56\pm 0.13) \times 10^{20}$ cm$^{-2}$ \\
$n_{e,0}$ & $9.6\times 10^{-3}$ cm$^{-3}$ \\
$\beta$ & 0.484 \\
$r_{c}$ & 0.047 Mpc \\
$f_{x [0.1,2.4]keV}$ (0.65 Mpc) & $6.5\times 10^{-12}$ erg\,s\per\,cm$^{-2}$ \\
$L_{x [0.1,2.4]keV}$ (1.3 Mpc) & $7.8 \times 10^{43}$ erg\,s\per\\
$M_{gas}$ (1.3 Mpc) & $3.3\times 10^{13}$ M$_{\odot}$ \\
$M_{tot}$ (1.3 Mpc) & $1.1\times 10^{14}$ M$_{\odot}$ \\
$M_{tot}/L_{B}$ (1.3 Mpc) & 77 \\
\hline
\end{tabular}
\end{center}
\end{table}

The cluster surface brightness was obtained by first subtracting the
point X-ray sources from the X-ray photon events file and producing a
sky subtracted azimuthally averaged count rate profile (see Figure 2).
Since the PSPC has a much lower background level in the hard band and
the cluster emission tends to dominate the hard band, we have chosen
only the data between channels 42 to 201 ($\sim$ 0.4-2 keV) to produce
the surface brightness profile.
\begin{figure}
\epsfxsize 230pts \epsfbox{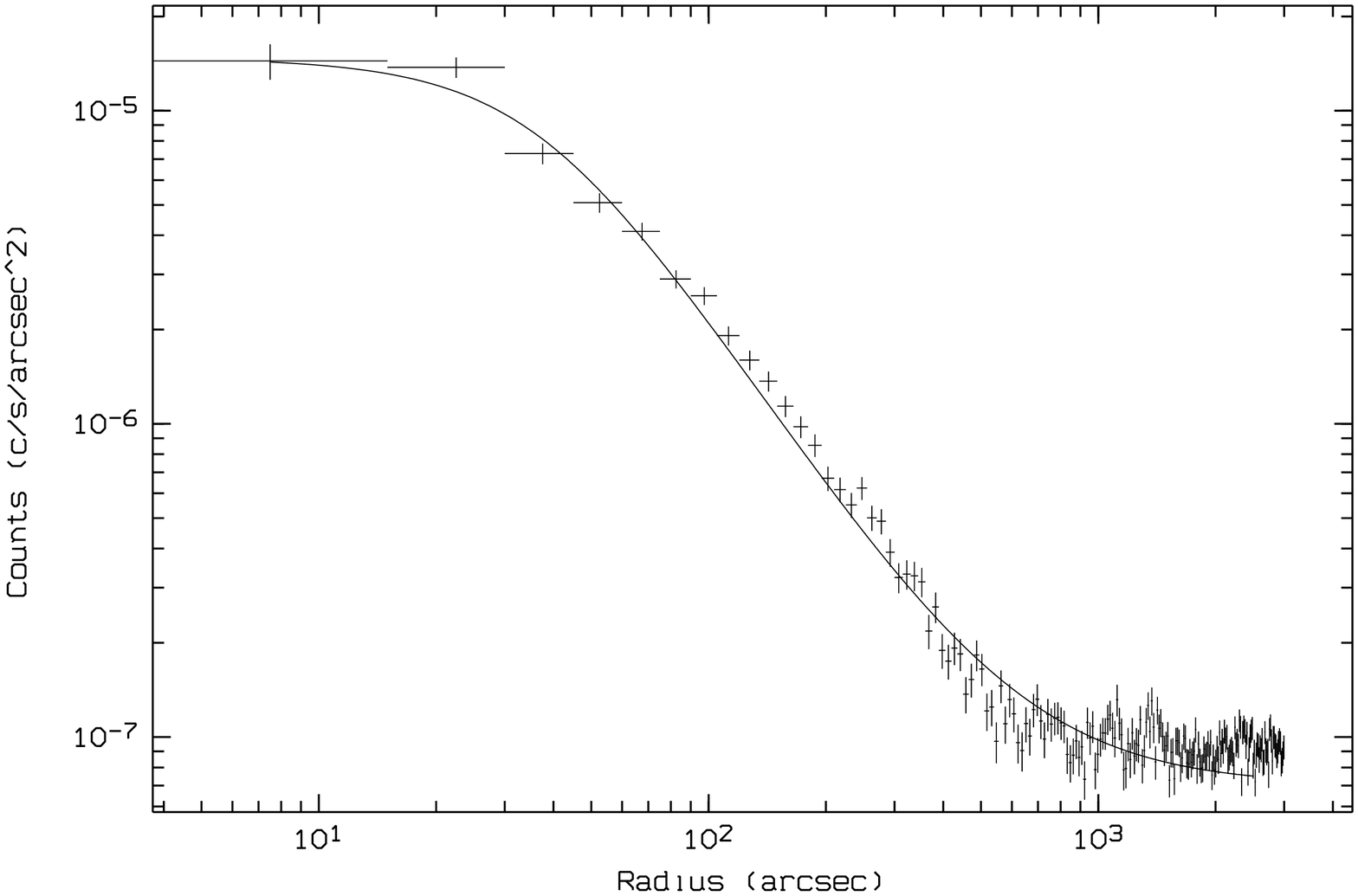} \label{f:sx}
\caption{The cluster X-ray surface brightness profile in the energy range 
of 0.4-2 kev. The data points with error bars are shown along with the
cluster model X-ray surface brightness profile (including background)
after convolution with the PSF (solid line).}
\end{figure}

The radially averaged X-ray surface brightness profile after the
deconvolution of the PSF, was well fitted by a function of the form :
\begin{equation}
S_{x}(r)=S_{x,0} [1+(r/r_{c})^2]^{-3\beta +1/2} \label{eq:sx}
\end{equation}
with $\beta = 0.484$ and core radius $r_{c} = 35.5^{''}$ or 0.047 Mpc.

\section{Mass distributions} \label{s:mass}
In this section we will described the method we used to deduce the
cluster total mass, gas mass and other associated physical parameters
from the X-ray data and compare some of these parameters
(e.g. velocity dispersion) with the optical data available.
We will assume that the gas is in hydrostatic equilibrium and isothermal.
The X-ray surface brightness given in equation~\ref{eq:sx} corresponds 
exactly to a gas distribution of 
\begin{equation}
\rho_{g}/\rho_{g,0} = [1+(\frac{r}{r_{c}})^{2}]^{-3\beta/2}
\label{eq:gas}
\end{equation}
if the gas extends all the way to infinity. Through the equation of
hydrostatic equilibrium, the gas distribution is related to the cluster 
gravitational potential $\phi$ as follows: 
\begin{equation}
\rho_{g}/\rho_{g,0} = \exp [\frac{\mu m_{p}}{k T_{g}} (\phi_{0}-\phi)]
\label{eq:rho}
\end{equation}
where $T_{g}$ is the gas temperature, $m_{p}$ is the proton mass and
$\mu$ is the mean molecular weight of the gas in proton mass units.
Thus the cluster gravitational potential corresponding to the gas
distribution of equation~\ref{eq:gas} is given by
\begin{equation}
\phi(r)-\phi_{0}=\frac{3\sigma^{2}_{0}}{2}\ln[1+(\frac{r}{r_{c}})^{2}].
\label{eq:pot}
\end{equation}
where $\sigma_{0}$ is the central 3-D velocity dispersion and is related to
$\beta$ through $\beta=\frac{\mu m_{p}\sigma_{0}^{2}}{kT_{g}}$.

Thus from Poisson's equation the total mass density is given by:
\begin{equation}
\rho_{tot} (r) = \frac{3\beta kT_{g}}{4\pi \mu m_{p} G r_{c}^{2}} \frac{3+(r/r_{c})^{2}}{[1+(r/r_{c})^{2}]^{2}}   \label{eq:bind}
\end{equation}
and the cluster total mass is given by:
\begin{equation}
M_{tot}(r)=\frac{3\beta kT_{g}r_{c}}{\mu m_{p}G} \frac{(r/r_{c})^{3}}{1+(r/r_{c})^2}
\end{equation}

From the X-ray spectral data, we found the best estimate emission
measure and temperature assuming Raymond-Smith spectra. Thus we have
the X-ray flux, $f_{x}$, within a radius of $500^{''}$ (i.e. 0.65 Mpc)
over the {\em Rosat} energy band of [0.1,2.4] keV, from the spectral
data. The X-ray surface bightness profile given in Figure 2 gives the
shape of the profile, namely $\beta$ and $r_{c}$.  The central
electron density $n_{e,0}$ can then be estimated from $f_{x}$, $\beta$
and $r_{c}$ (see Table~\ref{t:x2717}). The cluster total mass and gas
mass within the radius of $1000^{''}$ (or 1.3 Mpc), i.e. the maximum
extent where there is still detectable X-ray emission from the surface
brightness profile, are given in Table~\ref{t:x2717}. The radius of
$1000^{''}$ (or 1.3 Mpc) is chosen such that no extrapolation would be
necessary in calculating the masses from the X-ray data. The total
X-ray luminosity calculated up to the same radius of 1.3 Mpc was found
to be consistent with the $L_{x} - T_{g}$ relation given by Ebeling
(1993).

\begin{figure}
\label{f:galx}
\psfig{file=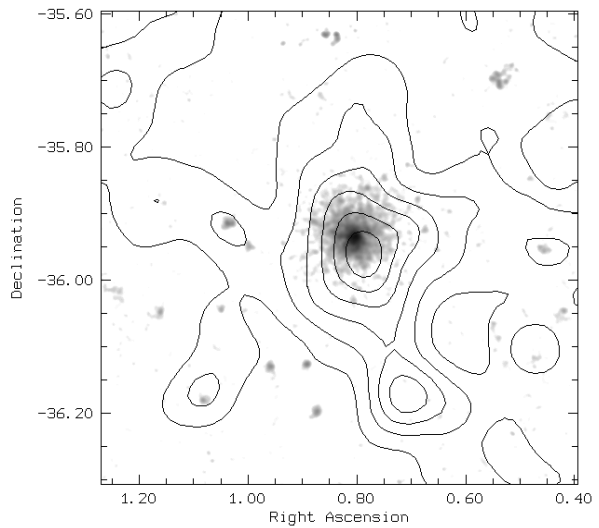,width= 8 cm}
\caption{ An image of the {\em Rosat} PSPC image superimposed on the galaxy 
number counts isocontours (obtained from Colless 1987). The axes are in decimal degrees. The contour levels
are (0.06, 0.12, 0.18, 0.24, 0.3, 0.36) galaxies\,arcmin\,$^{-2}$. }
\end{figure}

We compare the X-ray deduced velocity dispersion of the
cluster with the measured galaxy velocity dispersion. From Jean's
equation for a spherical, steady state system with isotropic velocity
distribution, we can deduce the 3-D velocity dispersion $\sigma$ from
the potential given in equation~\ref{eq:pot} and the total mass
density (equation~\ref{eq:bind}) if the galaxy number distribution
follows that of the total mass. Thus the 3-D velocity dispersion in this case 
is given by 
\begin{equation}
\sigma^{2}(y)=\frac{1+y^{2}/2}{1+y^{2}/3} \sigma_{0}^{2}
\end{equation}
where $y=r/r_{c}$ is the 3-D radius in core radius units and
 $\sigma_{0}$ is the 3-D velocity dispersion at the centre.  In
 producing the above analytic expression, we have assumed that the
 cluster extends to infinity and that the galaxies can be treated as
 test particles in the cluster potential with a distribution that
 follows the mass. Figure 3 shows the galaxy number count isocontours
 overlaid on the X-ray image.  The southern structure in the isopleth
 map was identified with a background group (Colless 1987). Note also
 the coincidence between the X-ray emission and the optical
 structures. Since the cluster is fairly regular, it can be circularly
 averaged to produce a galaxy number density distribution. We found
 that the galaxy density distribution thus deduced agreed well with
 the shape of the projected mass distribution, thus justifying the
 mass-follows-light assumption.  The line of sight velocity dispersion
 $\sigma_{los}$, can be projected from the 3-D dispersion through
 (Merritt 1987):
\begin{equation}
\Sigma(x) \sigma_{los}^{2} (x) = 2 \int^{\infty}_{x} \rho_{tot} \sigma^{2} (y) 
\frac{y dy}{\sqrt{y^2-x^2}}
\end{equation}
where $\Sigma$ is the projected mass density, $x=R/r_{c}$ is the projected 
radius in core radius units. Therefore, the line of sight velocity 
dispersion $\sigma_{los}$ is 
given by 
\begin{equation}
\sigma_{los}^{2}(x) =\frac{9}{8} [\frac{1+2x^{2}/3}{1+x^{2}/2}] \sigma_{0}^{2}
\end{equation}
(see Appendix of Mellier {\em et al} 1994, but note there is a
misprint). However, the measured velocity dispersion in most cases is the average of $\sigma_{los}$ within a certain radius. We derive the $\overline{\sigma_{los}}$ from the above and obtain 
\begin{equation}
\overline{\sigma_{los}}^{2} (<x) = \frac{3}{4} (2+\frac{1}{x^{2}}) \sigma_{0}^{2}
\end{equation}
We can thus estimate $\sigma_{0}^{2}$ from $\beta$ and
$T_{g}$ and deduce $\overline{\sigma_{los}}(<2$\,Mpc$)= 429\pm_{-22}^{+32}$ km\,s
\per. Note that $\overline{\sigma_{los}}^{2}(<2$\,Mpc$) = 1.5 \sigma_{0}^{2}$, thus
the difference is significant and the usual assumption of a constant
velocity dispersion with radius could easily produce the difference
between $\beta_{spec}$ and $\beta_{imag}$ noted in many clusters
(Sarazin 1988).  However, in the case of A2717, this difference is
still not large enough to obtain consistency.  The measured
$\overline{\sigma_{los}}$ of A2717 from the galaxies within a radius
of 2 Mpc was $\overline{\sigma_{los}}=547^{+92}_{-72}$ km\, s\per
(Colless {\em et al.}  1987). Thus the measured galaxy velocity
dispersion was slightly higher than that deduced from the X-ray
data. As mentioned in Colless {\em et al.} (1987), there is evidence
of foreground and background groups that may have contaminated the
measurements and thus overestimate the velocity dispersion.

Given the electron density and gas temperature we can calculate the central 
cooling time, $t_{cool}$, given by
\begin{equation} 
t_{cool} \sim 8.5\times 10^{10} (\frac{n_{e}}{10^{-3} cm^{-3}})^{-1} 
(\frac{T_{g}}{10^{8} K})^{0.5} yr
\end{equation} 
(Sarazin 1988) and for A2717, it amounts to $\sim 3.4$ Gyr which is
smaller than the Hubble time ($\sim 10$ Gyr). Thus in the absence of
heating, cooling occurs within the centre of the cluster with a
cooling radius of $\sim 90$ kpc. Pending further evidence of the
existence of a cooling flow, this maybe one of the few cases where a
WAT is found in a cooling flow cluster (c.f. Burns 1990).

\section{Radio Observations} \label{s:radio}
The WAT identified with the central D-galaxy was observed at the
Molonglo Observatory Synthesis Telescope (MOST) at 843 MHz with a
resolution of $43^{''}\times 73^{''}$ in 1988, at the VLA at 1.4 GHz
in the B/C configuration (resolution of $9^{''}\times 10^{''}$) and at
4.9 GHz in the C/D configuration (resolution of $9^{''}\times
10^{''}$) in 1989. The central parts of the source was also imaged at
the Australia Telescope at a high resolution ($2^{''}\times 4^{''}$) with a
6km array in 1992.  In this paper, we will concentrate on the VLA
images.

\subsection{ Analysis of Radio Data}
The 1.4 GHz image was calibrated using 3C48 as the primary calibrator,
but a different primary calibrator 3C286 was used for the 4.9 GHz
observation.  The images were processed using the standard AIPS
package for radio synthesis data. Figure 4 shows the 1.4 GHz image
superimposed on the X-ray image. The two tails are reasonably
symmetrically located around the centre of the cluster and are bending
southwards. The 4.9 GHz image smoothed to the same resolution as the
1.4 GHz image is shown in Figure 5. The eastern tail disappears below
the sensitivity limit in this image, due to the steepness of the
spectral index in the tail.  The spectral index steepens dramatically
from the flat core ($\alpha \sim -0.5$) to the lobes where $\alpha
\sim -3$. The linear size of the WAT was estimated from the 1.4 GHz
image to be $> 360$ kpc. The total radio luminosity of the WAT at 1.4
GHz is estimated to be $5.8\times10^{24}$ W Hz$^{-1}$, which is just
below the FR break. Such moderate radio powers are typical of radio
sources associated with cluster dominant galaxies (Heckman 1980) and
of medium sized WATs (Burns 1986).

A comparison of the redshifts of the D-galaxy and the cluster shows
that the D-galaxy is moving away from us at a speed of $\sim 300$ km
s$^{-1}$. In general the galaxies with
which the WATs are associated move through the intracluster medium
(ICM) very slowly, if at all, rarely obtaining velocities greater than
about a half their stellar dispersion (which corresponds to $\approx$
200\,km\,s\per\,) with respect to the cluster centre (Burns 1982,Eilek
1984,Bothun 1990). WATs were thought to be bent by the ICM through
their slow motion in the medium. However, recent detailed studies by
O'Donoghue {\em et al.}  (1993) have shown that this standard model
breaks down when examined in detail using the current theories of the
physics of the jets. Alternative models suggest that tails may be bent
by the bulk flow of velocities $\sim 1000$ km\per\,s in the ICM due to
subcluster mergings (Burns {\em et al.} 1986; O'Donoghue {\em et al.}
1993; Roettiger {\em et al.} 1993).

\begin{figure}
\label{f:rad21}
\psfig{file=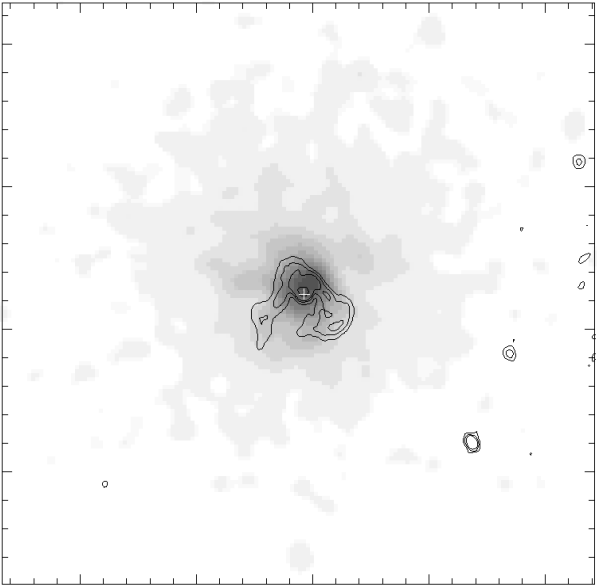,width= 8 cm}
\caption{Radio VLA image of the WAT at 1.4 GHz superimposed on the {\em Rosat} 
PSPC X-ray image in the hard band. The field size is $12^{'}\times
12^{'}$ with east to the left and north upwards. The radio
images is in contours of (0.8, 3.0, 7.0, 50.0) mJy\,beam\per. The
radio image was smoothed by a Gaussian to a resolution of FWHM
$11^{''}\times 11^{''}$ (c.f. Fig. 5).}
\end{figure}

\begin{figure}
\label{f:rad6}
\psfig{file=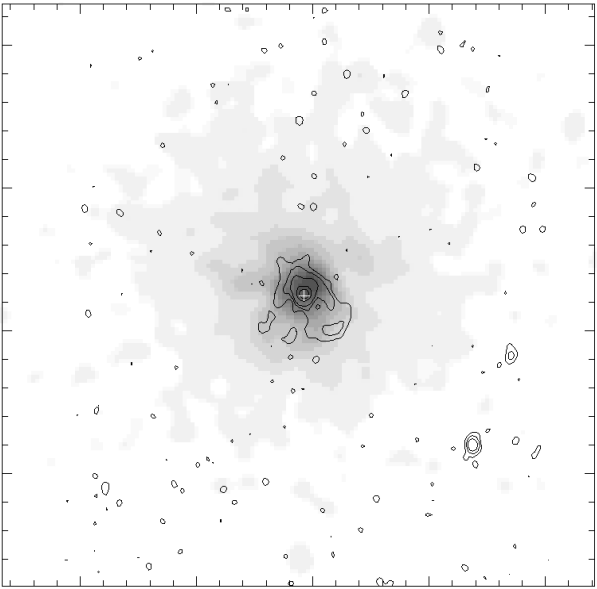,width= 8 cm}
\caption{A VLA contour image of the WAT at 4.9 GHz superimposed on a grey 
scale hard band PSPC image. The field size is $12^{'}\times 12^{'}$
with east to the left and north upwards. The radio contours are (0.07,
0.3, 0.8, 1.0, 2.0) mJy\,beam\per.  The radio image was smoothed by a
Gaussian to a resolution of FWHM $11^{''}\times 11^{''}$ (c.f. Fig. 4).}
\end{figure}

\section{Pressure balance} \label{s:press}
\subsection{Thermal pressure}
The thermal pressure due to the intracluster medium can be calculated
from the PSPC X-ray data by assuming that the gas is isothermal and
that the ion and electron temperatures are the same. Thus the thermal
pressure is approximated by $P_{th} \sim 2n_{e}kT_{g}$, where $T_{g}$
and $n_{e}$ are estimated from the X-ray surface brightness profile
and the X-ray spectra in section~\ref{s:xray}. Figure 6 shows the
thermal pressure as a function of radius.

\subsection{Minimum radiation pressure}
The radio data enable the minimum internal pressure of the tails to
be calculated (Burbidge 1958). The tails would be at a minimum
pressure when the equipartition condition is satisfied, namely the
energy density of the relativistic particles is equal to that of the
magnetic field (Pacholczyk 1970). Indeed given the thermal pressure of
the surronding medium, we can verify if the tails are in
equipartition.  We have adopted the method given by Killeen {\em et
al.} (1988) for calculating the minimum pressure. The following
assumption were made: 1) the jets are perpendicular to the line of
sight and the magnetic field lines are perpendicular to the line of
sight; 2) the energy ratio $k$ between relativistic electrons and
relativistic protons plus thermal particles is 1; 3) the source
spectra is a power law extending from 10 MHz to 100 GHz with a
spectral index of $\alpha=-2.3$; 4) the filling factor is 1. 
Amongst the above assumptions, the two most uncertain factors are $k$
and the filling factor. The parameter $k$ can range from 1 to 2000
(Pacholczyk 1970) and a factor of 100 in k produces roughly an order
of magnitude difference in $P_{min}$. The filling factor can easily be
$< 1$.  However, an increase in $k$ increases $P_{min}$ and a decrease
in the filling factor increases $P_{min}$. Thus by choosing $k=1$ and
filling factor $=1$, we have really calculated the minimum $P_{min}$.

\begin{figure}
\label{f:press}
\epsfxsize 250pts \epsfbox{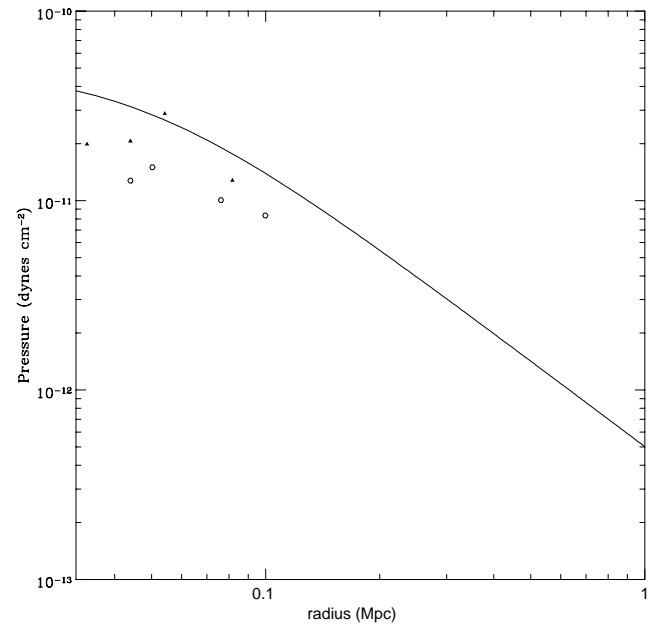}
\caption{A comparison of the thermal X-ray pressure (solid curve) with that 
of the minimum radiation pressure. The filled triangles show the minimum 
pressure for each slice of the western jet and the open circles are for that 
of the eastern jet.}
\end{figure}

In order to calculate the internal pressure exerted by the particles
in the radio source, the 1.4 GHz image was smoothed with a circular
beam of $11^{''}\times 11^{''}$ and Gaussians were fit to slices
perpendicular to the jet to estimate the peak flux density and size of
each slice. The minimum radiation pressure was calculated for each
slice using a program kindly supplied by Geoff Bicknell (see
Figure 6). An average spectral index of $\alpha \sim -2.3$
was used for each slice of the tails. Given the uncertainties
involved, Figure 6 shows that the minimum pressure deduced
from the radio data is in good agreement with the X-ray pressure,
i.e. within a factor of 2-3. This suggests that the tails are likely
to be in equipartition, if we assume that the tails are in pressure
equilibrium with the X-ray gas. Note that the minimum pressure plotted
are truly the lower limits, due to the choice of $k$ and filling
factor. Feretti {\em et al.} (1992) showed in a study of $\sim 40$
tailed radio sources, including WATs and NATs, that the thermal
pressure is always either comparable to or higher than the minimum
$P_{min}$. A study of 22 low radio luminosity sources by Morganti {\em
et al.} (1988) had found a similar result, suggesting that low
luminosity radio sources can be confined by thermal pressure alone.

\section{Conclusions}
We have presented the X-ray, radio data of the cluster A2717 and
analysed the data along with the optical data available in the
literature. The cluster has a total X-ray luminosity of $7.8\times
10^{43}$ erg\,s\per within a radius of 1.3 Mpc and a well determined
temperature of $1.9^{+0.3}_{-0.2}\times 10^{7}$ K. The X-ray
determined total mass was $\sim 1.1\times 10^{14}$ M$_{\odot}$ within
1.3 Mpc, comparable to the virial mass given all the
uncertainties. The gas mass was found to be $\sim 30$\% of the total
mass.

The thermal pressure from the ICM was found to be comparable to the
minimum pressure in the tails of the radio galaxy associated with the
central D galaxy, suggesting that the WAT is most likely to be in
equipartition with the surrounding media. The X-ray image of the
cluster gas at the PSPC resolution was almost structureless and no
obvious interaction between the WAT and the ICM was seen. To obtain
detailed morphological information that would show the possible
interaction between the radio galaxy and the X-ray medium, as seen in
the case of NGC 1275 (B\"ohringer {\em et al.}  1993), we need a
higher resolution X-ray image.

%\pagebreak

{\em Acknowledgements}: We would like to thank Monique Arnaud for
kindly providing the X-ray spectral analysis program, Geoff Bicknell
and Neil Killeen for their program to calculate the minimum pressure
and Matthew Colless for useful discussions on the optical data. We
acknowledge the use of the COSMOS/UKST Southern Sky Catalogue supplied
by the Anglo-Australian Observatory. HL acknowledges the support of
the Australia-France collaboration fund.

\end{document}